\documentclass[english,aps,preprint]{revtex4}
\usepackage{graphicx}
\usepackage{amssymb}

\makeatletter

\usepackage{babel}
\makeatother
\begin{document}

\preprint{BROWN-HET-1402}

\title{q-Deformed de Sitter/Conformal Field Theory Correspondence}

\author{David A. Lowe}

\email{lowe@brown.edu}

\affiliation{Physics Department, Brown University, Providence, RI 02912, USA}

\begin{abstract}
Unitary principal series representations of the conformal group appear
in the dS/CFT correspondence. These are infinite dimensional irreducible
representations, without highest weights. In earlier work of G\"uijosa
and the author it was shown for the case of two-dimensional de Sitter,
there was a natural $q$-deformation of the conformal group, with
$q$ a root of unity, where the unitary principal series representations
become finite-dimensional cyclic unitary representations. Formulating
a version of the dS/CFT correspondence using these representations
can lead to a description with a finite-dimensional Hilbert space
and unitary evolution. In the present work, we generalize to the case
of quantum-deformed three-dimensional de Sitter spacetime and compute
the entanglement entropy of a quantum field across the cosmological
horizon.
\end{abstract}
\maketitle

\section{Introduction}

Despite much progress in understanding string theory in backgrounds
with negative cosmological constant, the study of the more physically
relevant case of positive cosmological constant is still in its infancy.
One proposal for formulating quantum gravity in de Sitter space has
been proposed in \cite{Strominger:2001pn}. See also \cite{Park:1998yw,Park:1998qk,Bousso:2001mw,Spradlin:2001nb,Strominger:2001gp,Ghezelbash:2001vs,Li:2001pj,Balasubramanian:2001nb,Nojiri:2001mf,Klemm:2001ea,Das:2002he,Kabat:2002hj,Leblond:2002tf,Larsen:2002et,Balasubramanian:2002zh,Kristjansson:2002yb,Goldstein:2003qf,Larsen:2003pf,Astefanesei:2003gw}
for further developments and background. This proposal involves formulating
a holographic dual to de Sitter, with the isometry group of de Sitter
identified with the conformal group of the dual field theory.

At first sight this proposal seems to conflict with the idea that
de Sitter space should be described by a theory with a finite-dimensional
Hilbert space, bounded by $e^{S_{BH}}$, with $S_{BH}$ the Bekenstein-Hawking
entropy associated with the cosmological horizon \cite{Bousso:2000nf,Banks:2002wr}.
The unitary representations of the conformal group have continuous
unbounded energy spectra, so even the single particle Hilbert space
is infinite-dimensional. 

In recent works new versions of the dS/CFT correspondence have been
proposed built on finite-dimensional Hilbert spaces. In \cite{Parikh:2004ux}
it is suggested the Hilbert space be built out of fermionic degrees
of freedom which transform under the $SO(d-1)\times\mathbb{R}$ symmetry
that leaves the horizon of a comoving observer fixed. It is argued
a S-matrix invariant under the full de Sitter isometries may nevertheless
be constructed.

In the present paper we further elaborate on the alternative proposal
of \cite{Guijosa:2003ze} where the de Sitter isometry group is replaced
with a quantum deformed version. See \cite{Pouliot:2003vt,Jevicki:1998rr}
for earlier related work. In the original work \cite{Guijosa:2003ze}
the proposal was made in its most detailed form for the case of two-dimensional
de Sitter space, where the isometry group is $SL(2,\mathbb{R})$ which
is deformed to $SL_{q}(2,\mathbb{R})$ with $q=e^{2\pi i/N}$. Cyclic
unitary irreducible representations of the quantum group were found
to become unitary principal series representations of the conformal
group as $N\to\infty$. The construction agrees with the proposal
of Strominger in this limit, but can yield a finite-dimensional Hilbert
space for $N$ finite. These representations form natural building
blocks for a new formulation of dS/CFT that avoids problems associated
with non-unitarity and continuous spectra. One question left unanswered
in \cite{Guijosa:2003ze} was the detailed relation between $N$ and
the cosmological constant. This was related to the fact that the Bekenstein-Hawking
entropy formula is degenerate in two-dimensions, where the horizon
is a point. 

In the present work we generalize the construction of \cite{Guijosa:2003ze}
to three-dimensional de Sitter, with a view to obtaining a better
understanding of the Bekenstein-Hawking entropy in this framework.
We find it is necessary to modify the proposal for the inner product,
in order to obtain unitary representations of the quantum isometry
group. Nevertheless, these representations deform to unitary principal
series representations of the de Sitter isometry group $SL(2,\mathbb{C})$
in the $N\to\infty$ limit, where the construction coincides with
the original proposal of Strominger.

We perform the analog of the 't Hooft brick-wall entropy calculation
of the horizon entropy for the quantum deformed three-dimensional
de Sitter space. The result is regulated for finite $N$, and here
we are able to relate the cosmological constant to the parameters
that characterize the representations of the quantum isometry group.
In future work we hope to extend these results to a quantum deformation
of the full Virasoro algebra, which can potentially yield a complete
description of the entropy of de Sitter space.

\section{Basic Setup}

Three-dimensional de Sitter space can be viewed as a hyperboloid embedded
in flat Minkowski space $(t,x,y,z)$\[
-t^{2}+x^{2}+y^{2}+z^{2}=R^{2}~,\]
and so inherits the isometry group $SO(3,1)\approx SL(2,\mathcal{\mathbb{C}})$.
For generic masses larger than the Hubble scale, the representations
of the isometry group relevant for the dS/CFT correspondence are the
unitary principal series representations \cite{Guijosa:2003ze,Balasubramanian:2002zh}.
These representations are described in detail in \cite{Gelfand:1,Naimark:1}.
They can be realized as smooth functions on the complex plane with
asymptotic behavior\begin{equation}
f(z,\bar{z})|_{|z|\to\infty}\sim z^{2\tau}\bar{z}^{2\tilde{\tau}}~,\label{eq:falloff}\end{equation}
with $\tau=\frac{s}{4}+\frac{i\rho}{4}-\frac{1}{2}$ and $\tilde{\tau}=-\frac{s}{4}+\frac{i\rho}{4}-\frac{1}{2}$,
with $s$ integer and $\rho$ real, and which have finite norm with
respect to the inner product\begin{equation}
(f,g)=\int d^{2}z~\overline{g(z,\bar{z})}~f(z,\bar{z})~.\label{eq:psnorm}\end{equation}
This is the norm induced on the boundary by the standard Klein-Gordon
norm in the bulk \cite{Guijosa:2003ze}. A generator of $SL(2,\mathcal{\mathbb{C}})$
can be written

\begin{equation}
c_{n}L_{n}+\bar{c}_{n}\bar{L}_{n}~,\label{eq:sltcgens}\end{equation}
where $\bar{c}_{n}$ are the complex conjugates of the $c_{n}$ with
$n=-1,0,1$. The $L_{n}$ satisfy an $sl(2,\mathbb{R})$ algebra.
Likewise the $\bar{L}_{n}$ satisfy an $sl(2,\mathbb{R})$ algebra
and commute with the $L_{n}$. Acting on the function $f(z,\bar{z})$
the generators may be written\begin{eqnarray*}
L_{-1} & = & \partial_{z}~,\quad L_{0}=-z\partial_{z}+\tau~,\quad L_{1}=-z^{2}\partial_{z}+2\tau z\\
\bar{L}_{-1} & = & \partial_{\bar{z}}~,\quad\bar{L}_{0}=-\bar{z}\partial_{\bar{z}}+\tilde{\tau}~,\quad\bar{L}_{1}=-\bar{z}^{2}\partial_{z}+2\tilde{\tau}\bar{z}~.\end{eqnarray*}
It is important to note that basis functions that diagonalize $L_{0}$
and $\bar{L}_{0}$ such as $z^{\alpha}\bar{z}^{\beta}$ lie outside
the space of functions of finite norm. If this were not the case,
it would be easy to show the representation was not unitary, by considering
the norm of, for example, $(L_{-1}+\bar{L}_{-1})z^{\alpha}\bar{z}^{\beta}$.
However provided we restrict to functions of finite norm, the action
of the generators maintain this condition, and the principal series
representation is indeed unitary. 

According to the dS/CFT proposal, the parameters that appear in this
representation are related to the mass $m$ and spin $s$ of fields
in the de Sitter background. Usually only scalar fields ($s=0)$ are
considered, where \[
\rho=\sqrt{{m^{2}-1}}~.\]
The bulk scalar field is dual to an operator in the CFT with conformal
dimension\[
h_{-}=-2\tau-2\tilde{\tau}-1=1-i\rho~.\]
However there is an equivalence \cite{Gelfand:1,Klimyk:1991fg} of
the representations under $\tau\to-1-\tau$, $\tilde{\tau}\to-1-\tilde{\tau}$,
for generic values of $\rho$, so the same representation also describes
operators with conformal weight\[
h_{+}=1+i\rho~.\]

One obtains the bulk field configuration by convolving the boundary
field configuration on past asymptotic infinity of de Sitter with
the bulk to boundary propagator. Due to the equivalence relation,
only a single representation of the boundary theory is needed to describe
both these operators. The Bunch-Davies vacuum is expected to be the
stable vacuum of de Sitter \cite{Goldstein:2003qf}, which selects
a particular linear combination of these operators as possible sources
in the boundary theory. This corresponds to a specific choice of bulk
to boundary propagator. See \cite{Guijosa:2003ze} for further discussion
of these points. 

In terms of the canonical basis of $su(1,1)\approx sl(2,\mathbb{R})$
generators that we will use below, the $L_{n}$ generators are

\begin{eqnarray}
L_{0} & = & \frac{1}{2}\left(X_{+}+X_{-}\right)\label{eq:vgens}\\
L_{-1} & = & \frac{i}{2}\left(X_{+}-X_{-}-H\right)\nonumber \\
L_{1} & = & \frac{i}{2}\left(X_{-}-X_{+}-H\right)~.\nonumber \end{eqnarray}
The $X_{\pm}$, $H$ notation is typical in the literature on quantum
groups \cite{Chari:1994pz,Klimyk:1997eb}. The {*}-structure inherited
from (\ref{eq:psnorm}) acts via\begin{eqnarray}
X_{\pm} & \to & -\bar{X}_{\mp},\qquad H\to\bar{H}\label{eq:starone}\\
\bar{X}_{\pm} & \to & -X_{\mp},\qquad\bar{H}\to H\nonumber \end{eqnarray}
or equivalently\begin{eqnarray*}
L_{n} & \to & -\bar{L}_{n}\\
\bar{L}_{n} & \to & -L_{n}~.\end{eqnarray*}
Another {*}-structure that will play an important role in the following
acts without the left-right exchange as

\begin{eqnarray}
X_{\pm} & \to & -X_{\mp},\qquad H\to H\nonumber \\
\bar{X}_{\pm} & \to & -\bar{X}_{\mp},\qquad\bar{H}\to\bar{H}\label{eq:startwo}\end{eqnarray}
so that \begin{eqnarray*}
L_{n} & \to & -L_{n}\\
\bar{L}_{n} & \to & -\bar{L}_{n}~.\end{eqnarray*}
This is equivalent to changing the inner product from (\ref{eq:psnorm})
to

\begin{equation}
(f,g)=\int d^{2}z~\overline{g(Pz,P\bar{z})}~f(z,\bar{z})=\int d^{2}z~\overline{g(\bar{z},z)}~f(z,\bar{z})\label{eq:newnorm}\end{equation}
where we have defined the parity operation as $Pz=\bar{z}$. We will
discuss this inner product in more detail below.

\section{$q$-deformed $\mathrm{dS}_{3}/\mathrm{CFT}_{2}$\label{sec:-q--deformed}}

As we will see a generalization of \cite{Guijosa:2003ze} to this
case is not entirely straightforward, owing to the fact that the isometry
group $SO(3,1)$ (or more precisely its complexification) is not simple.
This means there is not a unique $q$-deformation, but rather multi-parameter
families of $q$-deformation \cite{Woronowicz:1}. We will consider
a quantum deformation of $sl(2,\mathcal{\mathbb{C}})$ that preserves
the left-right factorization of the group elements and replaces the
$sl(2,\mathbb{R})\times sl(2,\mathbb{R})$ algebras by $U_{q}(su(1,1)_{\mathbb{R}})\times U_{q}(su(1,1)_{\mathbb{R}})$,
using the same notation as \cite{Guijosa:2003ze}. We will explain
this in more detail momentarily\footnote{Since the quantum group is defined in terms of polynomials of algebra generators, we will ignore the global distinction between $SO(3,1)$ and $SL(2,\mathbb{C})$.}.
It is worth noting this is not the usual $q$-deformation that has
been considered in much of the literature on the the quantum deformed
Lorentz group , where also $q$ is usually taken to be real, rather
than a root of unity, and where the quantum deformation is constructed
to maintain an $su_{q}(2)$ subalgebra.

Our basic building block will be the quantum group $SL_{q}(2,\mathbb{C})$,
or more precisely, the universal enveloping algebra $U_{q}(sl(2,\mathbb{C}))$
with complex coefficients built out of generators $K$, $X_{+}$,
$X_{-}$ satisfying \cite{Klimyk:1997eb}

\begin{equation}
KK^{-1}=K^{-1}K=1,\qquad KX_{\pm}K^{-1}=q^{\pm2}X_{_{\pm}},\qquad[X_{+},X_{-}]=\frac{K-K^{-1}}{q-q^{-1}}~.\label{eq:qalg}\end{equation}
 The universal enveloping algebra consists of the space spanned by
the monomials \begin{equation}
(X_{+})^{n}K^{m}(X_{-})^{l}\label{eq:monomial}\end{equation}
 with $m\in\mathbb{Z}$ and $n,l$ non-negative integers. The generator
$K$ can be identified with $q^{H}$, but it is important to keep
in mind the enveloping algebra is defined in terms of products of
$K$ rather than $H$. The universal enveloping algebra is a Hopf
algebra with comultiplication $\Delta$ defined as\begin{eqnarray*}
\Delta(K) & = & K\otimes K\\
\Delta(X_{+}) & = & X_{+}\otimes K+1\otimes X_{+}\\
\Delta(X_{-}) & = & X_{-}\otimes1+K^{-1}\otimes X_{-}\end{eqnarray*}
and with antipode $S$ and counit $\varepsilon$ defined as\begin{eqnarray*}
S(K) & = & K^{-1},\qquad S(X_{+})=-X_{+}K^{-1},\qquad S(X_{-})=-KX_{-},\\
\varepsilon(K) & = & 1,\qquad\varepsilon(X_{\pm})=0.\end{eqnarray*}
For the special case that $q$ is a root of unity, we can extract
a real subalgebra of this Hopf algebra, that we denote $U_{q}(su(1,1)_{\mathbb{R}})$.
This is done by defining a map \begin{equation}
\theta(X^{\pm})=-X^{\pm},~\theta(K)=K^{-1},\label{eq:theta}\end{equation}
and showing that the restriction $a^{*}=\theta(a)$ where $a\in U_{q}(su(1,1))$
and {*} is defined in (\ref{eq:startwo}), is compatible with the
algebra and comultiplication structure. 

The full $q$-deformed $sl(2,\mathbb{C})$ is then defined as two
mutually commuting left-right copies of $U_{q}(su(1,1)_{\mathbb{R}})$,
with generators $X_{\pm}$, $K$ and $\bar{X}_{\pm}$, $\bar{K}$.
In the classical case, $sl(2,\mathbb{C})$ is generated by elements
of the form (\ref{eq:sltcgens}). The corresponding statement in the
$q$-deformed case is that we allow for elements $A$ of $U_{q}(su(1,1)_{\mathbb{R}})\times U_{q}(su(1,1)_{\mathbb{R}})$
with complex coefficients, but then we impose a restriction with respect
to the real form that acts on the generators as\[
\phi(L_{n})=\bar{L}_{n},\quad\phi(\bar{L}_{n})=L_{n}\]
together with complex conjugation of numerical coefficients, such
that $\phi(A)=A$. This is equivalent to the action\begin{eqnarray}
X_{\pm} & \to & \bar{X}_{\mp},\quad H\to-\bar{H},\quad K\to\bar{K}\label{eq:realform}\\
\bar{X}_{\pm} & \to & X_{\mp},\quad\bar{H}\to-H,\quad\bar{K}\to K~.\nonumber \end{eqnarray}
To be specific, this is an involutive morphism, and co-morphism (i.e.
the order of generators is not changed, and the order in the co-product
also remains unchanged). We denote the final algebra $U_{q}(SL(2,\mathbb{C})_{\mathbb{R}})$
where the $\mathbb{R}$ denotes this is a real algebra (i.e. one with
six independent generators, versus, say, $U_{q}(SL(2,\mathbb{C}))$
in standard notation in the quantum group literature \cite{Chari:1994pz,Klimyk:1997eb},
that has three independent generators with arbitrary complex coefficients).

In order for a representation to be unitary, any state acted on by
products of the generators (\ref{eq:sltcgens}) should be positive,
semi-definite, vanishing only for the trivial state. For $q$ a root
of unity, irreducible representations are finite dimensional \cite{Klimyk:1997eb},
so one can always pick an orthonormal basis where $K$ and $\bar{K}$
are diagonal. It is easy see that the state $(X_{+}+\bar{X}_{-})|\psi\rangle$
where $|\psi\rangle$ is any basis element, has vanishing norm with
respect to the {*}-structure (\ref{eq:starone}). Therefore there
are no nontrivial unitary representations of the $q$-deformed algebra
with respect to (\ref{eq:starone}). Instead we will construct a non-unitary
representation that goes over to a unitary principal series representation
at $q=1$. Furthermore we will see this non-unitary representation
is actually unitary with respect to the different {*}-structure (\ref{eq:startwo}). 

The irreps we have in mind are constructed from a left-right tensor
product of two cyclic representations of the type discussed in \cite{Guijosa:2003ze}
in the context of $dS_{2}$. For $q$ a root of unity (we will take
$q=e^{2\pi i/N}$ with $N$ odd) there exists a class of finite-dimensional
irreducible representations of the quantum group $U_{q}(SL(2,\mathbb{C}))$
that can be realized on the $N$-dimensional basis $|m\rangle$ with
$m=0,\cdots,N-1$ \cite{Klimyk:1997eb} and parametrized by the complex
numbers $a,b,\lambda$:\begin{eqnarray*}
K|m\rangle & = & q^{-2m}\lambda|m\rangle\\
X_{+}|m\rangle & = & \left(ab+\frac{q^{m}-q^{-m}}{q-q^{-1}}\frac{\lambda q^{1-m}-\lambda^{-1}q^{m-1}}{q-q^{-1}}\right)|m-1\rangle\\
X_{-}|m\rangle & = & |m+1\rangle\end{eqnarray*}
supplemented by the additional cyclic operations\begin{equation}
X_{+}|0\rangle=a|N-1\rangle~,\qquad X_{-}|N-1\rangle=b|0\rangle~,\label{eq:cyclicop}\end{equation}
and the consistency condition\begin{equation}
|b|^{2}=-\prod_{j=0}^{N-1}s(j)\label{eq:consist}\end{equation}
where $s(j)\equiv ab+\frac{q^{m}-q^{-m}}{q-q^{-1}}\frac{\lambda q^{1-m}-\lambda^{-1}q^{m-1}}{q-q^{-1}}$.
For $a,b\neq0$ there are no highest- or lowest-weight states and
the representation is called cyclic. As in \cite{Guijosa:2003ze}
we will be interested in the case where we set $\lambda=q^{N-1}$,
and let us set\[
ab=\tau(\tau+1)-\frac{1}{4}\left(N^{2}-1\right)\]
which fixes $a$ and $b$ up to a mutual phase, when combined with
(\ref{eq:consist}). In the limit $N\to\infty$, it was shown that
this becomes a principal series representation of $sl(2,\mathbb{R})$,
characterized by the complex parameter $\tau$.

We build our candidate representation of the full $U_{q}(sl(2,\mathbb{C})_{\mathbb{R}})$
using a left-right tensor product of two of these representations
with

\begin{eqnarray}
\tau & = & -\frac{1}{2}+\frac{i\rho}{4}+\frac{s}{4},\quad\tilde{\tau}=-\frac{1}{2}+\frac{i\rho}{4}-\frac{s}{4}\label{eq:taulab}\end{eqnarray}
with $\rho$ real and $s$ integer as above. To check unitarity, it
suffices to check that $(c_{n}L_{n}+\bar{c}_{n}\bar{L}_{n})|\psi\rangle$
has positive norm for all choices of the complex coefficients $c_{n}$
with respect to the {*}-structure (\ref{eq:startwo}). Converting
to $U_{q}(su(1,1)_{\mathbb{R}})$ generators using (\ref{eq:vgens}),
we must check that $(\bar{c}_{0}H-c_{0}\bar{H})(c_{0}H-\bar{c}_{0}\bar{H})\geq0$,
$-(\bar{c}_{+}X_{-}+c_{+}\bar{X}_{+})(c_{+}X_{+}+\bar{c}_{+}\bar{X}_{-})>0$,
and $-(\bar{c}_{-}X_{+}+c_{-}\bar{X}_{-})(c_{-}X_{-}+\bar{c}_{-}\bar{X}_{+})>0$. 

Using the fact that $m$ and $\bar{n}$ are real, the first inequality
implies\[
\langle m,\bar{n}|(\bar{c}_{0}H-c_{0}\bar{H})(c_{0}H-\bar{c}_{0}\bar{H})|m,\bar{n}\rangle=|c_{0}m-\bar{c}_{0}\bar{n}|^{2}\langle m,\bar{n}|m,\bar{n}\rangle\geq0\]
which is trivially satisfied. Next we consider the second inequality,
which reduces to\[
|c_{+}|^{2}\langle m,\bar{n}|-X_{-}X_{+}-\bar{X}_{+}\bar{X}_{-}|m,\bar{n}\rangle>0~.\]
 The results of \cite{Guijosa:2003ze} can then be used to reduce
this expression to the inequality\[
-\tau(\tau+1)-\tilde{\tau}(\tilde{\tau}+1)>0\]
which holds assuming the form (\ref{eq:taulab}). No new conditions
emerge from considering the final inequality. 

In the limit $N\to\infty$, the representation becomes a principal
series representation of the individual left or right $sl(2,\mathbb{R})$
factors. Taking into account the elements of the $sl(2,\mathbb{C})$
algebra are built out of these $sl(2,\mathbb{R})$ generators with
complex coefficients (\ref{eq:sltcgens}), we obtain a principal series
representation of $sl(2,\mathbb{C})$ characterized by $(\rho,s)$
as described in the previous section. This is unitary with respect
to the norm (\ref{eq:psnorm}) at $q=1$, but non-unitary for $q\neq1$.
On the other hand, with respect to the new norm appropriate for the
{*}-structure (\ref{eq:startwo}) the representation is unitary for
general $q$ a root of unity. 

We therefore conjecture this new norm emerges from a non-perturbative
formulation of gravity in a de Sitter background. This bears some
similarity to Witten's inner product \cite{Witten:2001kn}, which
instead introduces the operation of $CPT$, and generates the adjoint
map $L_{n}\to L_{-n}$ typical of conformal field theory. Following
similar logic as \cite{Witten:2001kn} we conjecture the inner product
(\ref{eq:newnorm}) gives the appropriate unitary norm for $q=e^{2\pi i/N},q\neq1$.
Having to make this different choice for the inner product in the
quantum version of de Sitter, versus the classical $q=1$ case, is
not entirely unreasonable, since we expect the semiclassical limit
of quantum de Sitter to be rather subtle. In particular, the finite
$N$ theory includes modes that become non-normalizable as $N\to\infty$,
so it is not surprising the standard norm (\ref{eq:psnorm}) is non-unitary
at finite $N$. Conversely, while the new inner product is unitary
at finite $N$, it is no longer unitary at $N=\infty$, where it takes
the form (\ref{eq:newnorm}), assuming we impose the standard finiteness
and fall-off conditions (\ref{eq:falloff}). 

However we expect that with a proper definition of the bulk to boundary
map at finite $N$, it should be possible to see the new norm reproduces
ordinary perturbative quantum field theory in a de Sitter background
as $N\to\infty$. In effect, the extra parity operation can be undone
in the formulation of the finite $N$ bulk to boundary map. We hope
to explore this further in future work, however it is first necessary
to develop further $q$-deformed de Sitter geometry, and the harmonic
analysis on this space. This has not been studied for $q$ a root
of unity, however for $q$ real substantial progress has been made
in \cite{Olshanetsky}. There a $q$-deformed version of three-dimensional
Lobachevsky space is described, and analogs of the Bessel functions
and Macdonald functions are defined. We have checked that the quantum
Lobachevsky space can be analytically continued to describe a version
of $q$-de Sitter space at $q$ a root of unity, however further generalization
is necessary to obtain coordinates that yield the cyclic representations
of the quantum de Sitter group needed in the present context. We can
at least offer a more precise description of the geometry of the quantum
plane where the holographic CFT lives. We present this in the appendix
and will use these results in the following section when we compute
the entanglement entropy on de Sitter. 

In \cite{Guijosa:2003ze} it was pointed out that if one changes basis
so that $L_{0}$ is diagonal rather than $K$, then the $N\to\infty$
limit does not smoothly match that of the principal series representation
of $SL(2,\mathbb{R})$, where $L_{0}$ has continuous unbounded imaginary
eigenvalues. Instead one has a spectrum where the eigenvalues of $L_{0}$
range over approximately $-iN/2$ to $iN/2$ with imaginary spacings
of order $1$. This is also an issue for the situation we have described
above. We emphasize, in the basis where $K$ and $\bar{K}$ are diagonal
that limit $N\to\infty$ does correctly reproduce the principal series
representation, with each weight appearing with unit degeneracy as
required. This is one of many subtleties dealing with infinite-dimensional
representations. For the sake of definiteness we will refer to these
representations as Type I.

\begin{figure}
\includegraphics[%
  clip,
  scale=0.75]{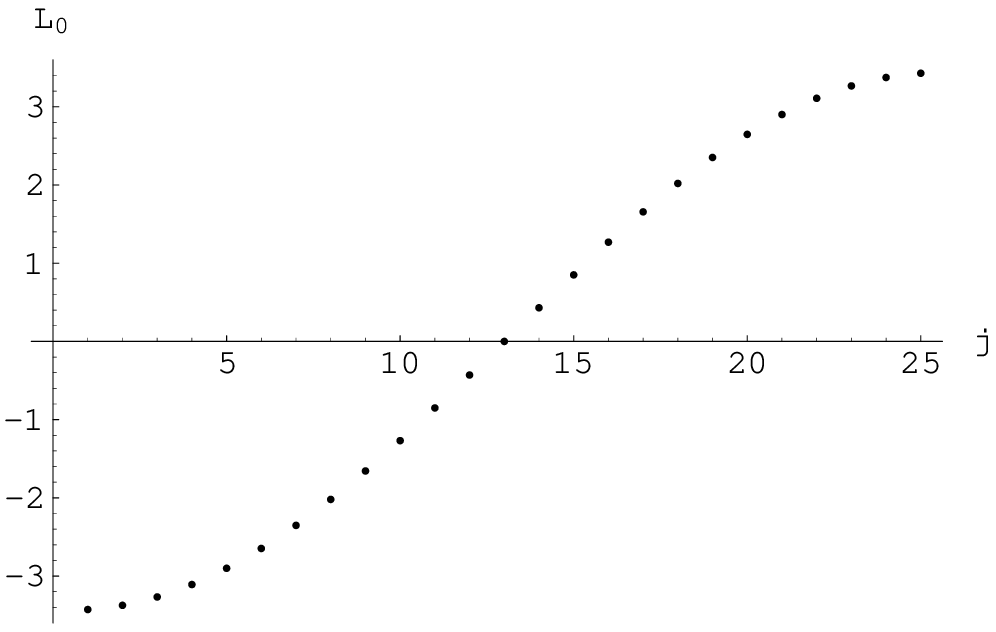}\includegraphics[%
  clip,
  scale=0.75]{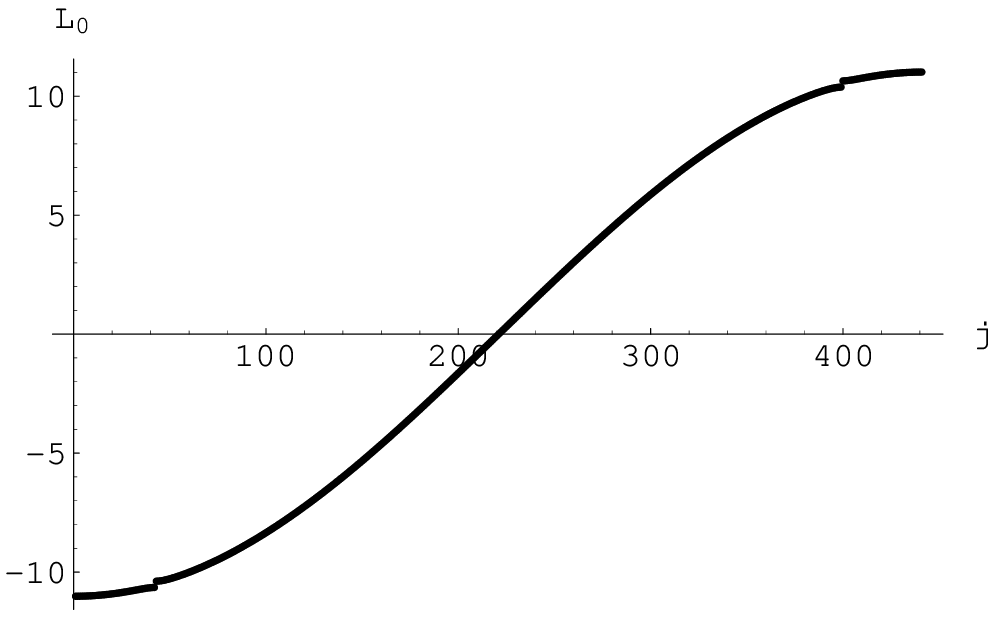}

Spectrum of $L_{0}$ in the reducible representation composed of a
direct sum of irreducible representations with $a_{k}=|a|e^{2\pi ik/N}$
for $k=0,\cdots,N-1$. The values $N=5$, and $N=21$ are shown. The
index $j$ labels the $N^{2}$ different eigenvalues in the reducible
representation.

\caption{\label{cap:specfig}}
\end{figure}

It is worth pointing out that by including $M^{2}$ different representations
of $U_{q}(sl(2,\mathbb{C})\mathbb{_{R}})$ it is possible to obtain
the principal series representation as $N\to\infty$ even in the basis
where $L_{0}$ and $\bar{L}_{0}$ are diagonal. The point is that
the eigenvalues of $L_{0}$ depend on the phase of $a$ (and likewise
for the right-moving factor). If we take a direct sum of $M$ different
irreducible representations $a_{k}=|a|e^{2\pi ik/M}$, with $k=0,\cdots,M-1$
we obtain a spectrum for $L_{0}$ with typical spacing of order $i/M$.
Combining this with the right moving factor, we can then build a reducible
representation of $U_{q}(sl(2,\mathbb{C})\mathbb{_{R}})$ built out
of a direct sum of $M^{2}$ different irreducible $(a_{k},b_{k};\tilde{a}_{l},\tilde{b}_{l})$
representations with $k,l=0,\cdots,M-1$, which has a continuum spectrum
for $L_{0}$ and $\bar{L}_{0}$ as $N\to\infty,$ provided we also
send $M\to\infty$. The simplest solution is to set $M=N$, but other
versions of the limit may also be worth exploring. Figure \ref{cap:specfig}
illustrates the approach to a continuous spectrum as $N=M$ is increased%
\footnote{Small gaps can appear in the spectrum as illustrated in the  $N=21$
case of figure \ref{cap:specfig}.%
} . We will refer to this class of representations as Type II.%

Let us comment on some of the physical characteristics of these representations.
As discussed in \cite{Balasubramanian:2002zh,Guijosa:2003ze}, $(L_{0}+\bar{L}_{0})/l$
generates time translations for a static patch observer. Here we have
restored factors of the Hubble radius $l$. For classical de Sitter
space with Lorentzian signature, the spectrum of this generator is
continuous and unbounded above and below. 

However the Type I $q$-deformed representations contain built in
infrared and ultraviolet cutoffs, $E_{IR}=1/l$, $E_{UV}=N/l$. It
is natural to identify the ultraviolet cutoff with the Planck scale
(for want of any other interesting UV scales in the problem), which
provides us with a way to fix $N=m_{pl}l$. Since the infrared cutoff
is fixed in terms of the Hubble scale, this points to a new discreteness
in the energy levels of quantum gravity in de Sitter space. It is
interesting to note $E_{IR}=10^{-34}eV$, too small to be accessible
in present day experiments. But the idea that quantum gravity could
be tested by performing ultra-low temperature experiments is intriguing.
It would be very interesting to develop other experimental tests of
the $q$-deformation model of de Sitter.

Type II representations, on the other hand, contain cutoffs of the
form $E_{IR}=1/Nl$, and $E_{UV}=N/l$. For large $N$, these representations
approach the continuous classical spectrum.

\section{Entanglement entropy}

't Hooft introduced the brick wall entropy model \cite{'tHooft:1985re}
to provide a simple toy model that naturally explains why the quantum
properties of a black hole are associated with its horizon. In this
model one places an ultraviolet cutoff on a field such that it vanishes
inside some fixed radial distance from the horizon. This cutoff explicitly
breaks general coordinate invariance. It is also necessary to introduce
an infrared cutoff to regulate the usual volume divergence of the
entropy associated with a field at finite temperature. 't Hooft noticed
that as one removes the ultraviolet cutoff, a divergence appears proportional
to the horizon area. The entire horizon entropy of the black hole
can then be accounted for by adjusting the ultraviolet cutoff and
the number of species of field. Subsequent work using analogous calculations
in flat spacetime \cite{Bombelli:1986rw,Srednicki:1993im,Kabat:1994vj,Callan:1994py,Holzhey:1994we}
emphasized the interpretation as entanglement entropy between degrees
of freedom inside and outside the horizon. The brick wall model has
been generalized to three-dimensional de Sitter space in \cite{Kim:1998zs}. 

The free energy depends on the statistics of particles under consideration.
The results for the free energy are the same up to coefficients of
order unity which we will neglect in the following. The free energy
is then given by\begin{eqnarray}
\beta F & = & -\int dn_{r}~dm~e^{-\beta E}\label{eq:freecl}\\
 & = & -\beta\int dE~dm~n_{r}(E,m)e^{-\beta E}\end{eqnarray}
where $E$ is the energy in static patch coordinates, $m$ is the
azimuthal quantum number, and $n_{r}(E,m)$ radial quantum number
fixed by\begin{equation}
n_{r}=\frac{1}{\pi}\int_{L}^{l-\epsilon}dr\frac{1}{g(r)}\sqrt{{E^{2}-\frac{m^{2}g(r)}{r^{2}}}},\qquad g(r)=1-\frac{r^{2}}{l^{2}}\label{eq:radnum}\end{equation}
for fixed brick-wall cutoff $\epsilon$ and infrared cutoff $L$.
We take units where Newton's constant $G_{N}=1/8$, and the cosmological
constant is $\Lambda=1/l^{2}$. In the second line of (\ref{eq:freecl})
we have integrated by parts. Usually the analysis proceeds by integrating
over $m$, subject to the condition that the square root in (\ref{eq:radnum})
remain positive. Then one is left with an $r$ integral that diverges
as $\epsilon\to0$, indicating that the states in question are localized
near the horizon. 

We can write (\ref{eq:freecl}) in a more group theoretic way as

\begin{equation}
\beta F=-\mathrm{Tr}~e^{-\beta E}\label{eq:ftrace}\end{equation}
where the trace is over states that live in the tensor product of
the isometry group $SL(2,\mathbb{C})$ with $\mathcal{M}$, a spacelike
hypersurface through the static patch. The $SL(2,\mathbb{C})$ representation
is the unitary principal series, and the trace is restricted to states
with positive energy and positive angular momentum, with the brick-wall
cutoff left implicit. As shown in \cite{Bousso:2001mw}, the expression
(\ref{eq:ftrace}) arises by taking the Euclidean/Bunch-Davies vacuum,
decomposing modes in global coordinates into those on two copies of
the static patch, and tracing over modes in the unobserved static
patch. 

We propose to generalize the expression (\ref{eq:ftrace}) to the
$q$-deformed case, where the brick-wall cutoff is no longer necessary.
Our main motivation for doing this is to illustrate the regularizing
properties of the $q$-deformed theory, in a simple toy model calculation.
The full formulation of the second quantized field theory in $q$-deformed
de Sitter will be left for future work. We expect the details of this
construction, and the choice of statistics of the fields will change
the free energy by coefficients of order unity. 

The first step is to perform the analog of the integration over the
spacelike surface $\mathcal{M}$. This should be generated by picking
a point in the quantum plane, described in the appendix, and summing
over distinct points obtained by acting with the quantum de Sitter
isometry group. As discussed in the appendix, this leads to a factor
$N^{2}$. The remaining trace over $sl_{q}(2,\mathbb{C})$ can the
be directly evaluated using the results of the previous section.

For the Type I (irreducible) representations we find an entropy $S_{\beta}$
that is dominated by the $L_{0}=0$, $\bar{L}_{0}=0$ term for $\beta\gtrsim2\pi l$.
This leads to an entropy approximately equal to \[
S_{\infty}=N^{2}~.\]
For the high temperature limit $\beta\to0$, where the entropy simply
counts the logarithm of the dimension of the entire Hilbert space,\[
S_{0}=N^{2}(N+1)^{2}/4~.\]
The simplest picture is one where this entropy $S_{0}$ is identified
with the horizon entropy of de Sitter $S_{\Lambda}=\pi/2G_{N}\sqrt{{\Lambda}}$
\cite{Bousso:2000nf}, allowing $N$ to be fixed in terms of $G_{N}\sqrt{\Lambda}$.
Alternatively, one can identify the entropy $S_{\beta=2\pi l}$ with
$S_{\Lambda}$. At this level of analysis, there seems to be no preference
to one scheme over the other. In fact, both are at odds with the identification
of the Planck scale with the ultraviolet cutoff that appears in the
representation of the quantum group, discussed near the end of section
\ref{sec:-q--deformed}. Of course, given the crudeness of the identification
of entanglement entropy with the gravitational entropy of the de Sitter
horizon, this mismatch is not surprising. Our hope is a more refined
stringy version of this entropy calculation will improve the situation.

For completeness, let us give the analogous results for entropy $S_{\beta}$
of the Type II (reducible) representations. As expected in the low
temperature limit, \[
S_{0}=N^{2}\]
 due to the sum over points on the spacelike hypersurface. At high
temperatures, we simply find the logarithm of the dimension of the
Hilbert space, \[
S_{\infty}=N^{2}(N^{2}+1)^{2}/4~.\]
On the other hand, for $\beta$ of order the Hawking inverse temperature
$2\pi l$, there is now a much smoother transition between the two
behaviors, since now many modes are relevant in the partition function.

This example of the analog brick wall calculation shows the $q$-deformation
does provide a regulator for the entropy in the quantum deformed de
Sitter spacetime. If this emerges as an effective description of string
theory in a de Sitter background, we can expect a full $q$-deformed
Virasoro symmetry plays a role. We hope to explore this in future
work.

\begin{acknowledgments}
The author thanks Alberto G\"uijosa, Daniel Kabat and Maulik Parikh
for valuable discussions. This work is supported in part by DOE grant
DE-FE0291ER40688-Task A, NSF U.S.-Mexico Cooperative Research grant
\#0334379, and US-Israel Binational Science Foundation grant \#2000359.
The author thanks ICTP, Trieste, Italy for hospitality during the
completion of this work.
\end{acknowledgments}
\appendix*

\section{Geometry of quantum plane}

This is described in \cite{Biedenharn}, for the case of $U_{q}(sl(2,\mathbb{C}))$,
and we adapt that presentation to the case of interest here. We introduce
two complex variables $z$ and $\tilde{z}$ which commute, and satisfy
the relations\[
z^{N}\sim1,\quad\tilde{z}^{N}\sim1\]
 together with a reality condition that sets the complex conjugate
of $z$ to $\tilde{z}$. These variables generate a cyclic representation
of $U_{q}(SL(2,\mathbb{C})_{\mathbb{R}})$ where we identify the generators
as\[
X_{+}=z^{-1}[z\partial_{z}+\alpha]_{q},\quad X_{-}=z[\alpha-1-z\partial_{z}]_{q},\quad K=q^{N-1-z\partial_{z}}~.\]
Here we introduced the symbol $[x]_{q}=(q^{x}-q^{-x})/(q-q^{-1})$,
and define $\alpha$ via $ab=\tau(\tau+1)-\frac{1}{4}\left(N^{2}-1\right)=[\alpha]_{q}^{2}$.
Similar definitions hold for the $\bar{X}_{\pm},\bar{H}$, with the
replacement $z\to\tilde{z}$ and $\alpha\to\tilde{\alpha}$, with
$[\tilde{\alpha}]_{q}^{2}=\tilde{\tau}(\tilde{\tau}+1)-\frac{1}{4}\left(N^{2}-1\right)$.
The quantum isometry algebra acts on the $N^{2}$-dimensional representation
generated by $z$ and $\tilde{z}$. 

A notion of $q$-derivative may be defined as\[
D_{q}f(z)=\frac{f(qz)-f(q^{-1}z)}{z(q-q^{-1})}=z^{-1}[z\partial_{z}]_{q}f(z)~.\]
With this definition $D_{q}z^{k}=[k]_{q}z^{k-1}$. Likewise a notion
of integration may be defined. This can be regarded as an analog of
contour integration\[
\int dz_{q}f(z)=z(q-q^{-1})\sum_{j=0}^{N-1}f(q^{j}z)\]
and is non-vanishing only for terms proportional to $z^{-1}$.

We interpret the quantum plane as a spacelike hypersurface through
quantum de Sitter space, which can be thought of as $N^{2}$ points.
To see this we consider the orbit of a point under the analog of the
two generators of translations in the plane, $X_{+}+\bar{X}_{+}$
and $X_{-}+\bar{X}_{-}$. Each of these generators has $N$ distinct
eigenvalues, and the individual terms satisfy $X_{\pm}^{N}\propto1$,
$\bar{X}_{\pm}^{N}\propto1$. Arbitrary products of these generators
span $N^{2}$ linearly independent terms. This corresponds to the
statement that isometries acting on a single point can generate a
spacelike hypersurface of de Sitter.

\bibliographystyle{/home/lowe/tex/brownphys}
\bibliography{/home/lowe/tex/qdscft}

\end{document}